\documentclass[twocolumn,aps,superscriptaddress,prl]{revtex4}
\usepackage{graphicx}
\usepackage{times}
\usepackage{amsmath}
\usepackage{amssymb}
\usepackage{units}
\usepackage{stmaryrd} 

\renewcommand{\vec}[1]{\boldsymbol{#1}}

\begin{document}
\preprint{0}

\title{Anisotropy effects on Rashba and topological insulator spin polarized surface states: a unified phenomenological description}

\author{Emmanouil Frantzeskakis}
\thanks{Current address: Synchrotron SOLEIL, L'Orme des Merisiers, Saint Aubin-BP 48, 91192 Gif sur Yvette Cedex, France}
\email{frantzeskakis@synchrotron-soleil.fr} 
\address{Institut de Physique de la
Mati\`{e}re Condens{\'e}e, Ecole Polytechnique
F{\'e}d{\'e}rale de Lausanne (EPFL), CH-1015 Lausanne,
Switzerland}

\author{Marco Grioni}
\address{Institut
de Physique de la Mati\`{e}re Condens{\'e}e, Ecole
Polytechnique F{\'e}d{\'e}rale de Lausanne (EPFL),
CH-1015 Lausanne, Switzerland}

\date{\today}

\begin{abstract}
Spin polarized two-dimensional electronic states have been observed in metallic surface alloys with giant Rashba splitting and at the surface of topological insulators. We study the surface 
band structure of these systems, in a unified manner, by exploiting recent results of $\mathbf{k}\cdot \mathbf{p}$ theory. 
The model suggests a different way to address the effect of anisotropy in Rashba systems. Changes in the surface band structure of various Rashba compounds can be captured by a single effective parameter
which quantifies the competition between the Rashba effect and the hexagonal warping of the constant energy contours.  
The same model provides a unified phenomenological description of the surface states belonging to materials with topologically trivial and non-trivial band structures.
\end{abstract}

\maketitle

\section{I. Introduction}

The spin degeneracy of electronic states can be lifted at surfaces
or interfaces by the combined effect of the spin-orbit (SO)
interaction and an out-of-plane electric potential gradient \cite{Bychkov1984}. The latter appears as an effective magnetic
field in the rest frame of the electrons giving rise to the
Rashba-Bychkov (RB) effect. Spin polarized low dimensional states have been
experimentally observed in the surface states of metals with a large
atomic SO coupling \cite{LaShell1996,Cercellier2006,Koroteev2004,Reinert2003}.
Such 2D electron gases (2DEG) can be described by a nearly-free electron (NFE) model where the RB SO interaction enters
as a $\mathbf{k}$-linear correction term in the effective Hamiltonian:
\begin{equation}
\textmd{H}_{\textmd{NFE-SO}}=b(k_{\textmd{x}}\boldsymbol\sigma_{\textmd{y}}-k_{\textmd{y}}\boldsymbol\sigma_{\textmd{x}})+\frac{k^{2}}{2m_{\textmd{s}}}\label{EQ1}
\end{equation}
where $b$ is the Rashba parameter, $m_{\textmd{s}}$ 
a scaled effective electron mass (i.e. $m_{\textmd{s}}=m^{*}/\hbar^{2}$) and
$\boldsymbol\sigma_{\textmd{i}}$ the Pauli spin matrices.

Similar spin polarized 2D states have been observed in a new class of quantum matter: the 3D 
topological insulators (TI) \cite{HsiehNatLett2008,Xia2009,Chen2009,HsiehNatLett2009,HsiehScience2009,HsiehPRL2009}. TI 
and RB systems are characterized by a similar momentum-dependent helical spin arrangement \cite{Xu2010}. 
Not surprisingly, H$_{\textmd{NFE-SO}}$ as introduced in Eq. (\ref{EQ1}), has been used as an approximation to describe the
low-energy, long-wavelength, surface band structure of the 3D TI materials \cite{Zhang2009,Wang2010}. In this case, the linear term is dominant and
$b=\hbar$v$_{\textmd{F}}$, where v$_{\textmd{F}}$ is the Fermi velocity at the Dirac point. The spin polarized surface states described by a 
TI effective Hamiltonian must reflect the topologically non-trivial band structure of the bulk. The RB surface states have a similar origin in the SO interaction, but correspond to 
trivial bulk topologies and always give rise to pairs of constant energy contours encircling the spin degenerate points \cite{FuPRB2007,Xia2009}.
Despite these differences, H$_{\textmd{RB}}$ (for RB systems) and H$_{\textmd{TI}}$ (for TI materials) are essentially 
identical. 

Greatly different values of $m_{\textmd{s}}$, and hence of the quadratic term in Eq. (\ref{EQ1}) cannot be the only explanation for the topological differences 
reflected in the surface states of the two kinds of systems. 
For instance, the overall band structure of RB-split Pb quantum well states with very heavy effective masses \cite{Dil2008} is clearly different from that of 3D 
TI materials \cite{HsiehNatLett2008,Xia2009,Chen2009,HsiehNatLett2009}. Even more interestingly, the surface state of Bi$_{2}$Se$_{3}$ has been fitted with 
an effective mass of only $0.18m_{\textmd{e}}$, a value lower than those encountered in many RB systems \cite{Wang2010}.

Here we investigate which parameters in H$_{\textmd{NFE-SO}}$ determine the topological differences reflected
in the surface states of RB and TI materials with the same structural symmetry. To this end, anisotropy effects
are very critical and arise naturally when the isotropic RB Hamiltonian [Eq. (\ref{EQ1})] is modified to describe real materials. The Rashba parameter in Eq. (\ref{EQ1})
was initially associated with the surface potential gradient \cite{Bychkov1984}, but large quantitative discrepancies in the size and anisotropy of the band splitting were observed because
H$_{\textmd{NFE-SO}}$ disregards the important influence of the ion cores \cite{LaShell1996}. One must therefore take into account the atomic structure and the corresponding
electronic charge distribution, either explicitly by using a site-spin representation
\cite{Liu2009,Liu22009,Frantzeskakis2010}, or implicitly by
modifying the effective H$_{\textmd{NFE-SO}}$. A single parameter cannot capture the
anisotropic band structure and different approaches have
been proposed. DFT calculations have stressed the importance of the wavefunction
asymmetry near the ion cores \cite{Bihlmayer2006,SakamotoTl2009}, while an independent NFE approach enhanced the standard model
with an \textit{ad hoc} in-plane potential gradient \cite{Premper2007}. 

Recent data on Bi- and Pb-based surface alloys on metallic and semiconducting substrates have provided an experimental 
counterpart to the theoretical studies  \cite{Ast2007,Pacile2006,Frantzeskakis2008,He2008,Gierz2009,FrantzeskakisCrep2009}.
Those \textit{giant} RB systems exhibit both a very large RB splitting and clear anisotropies. The anisotropy manifests itself in the shape of the constant-energy (CE)
contours and the existence of an out-of-plane component of the spin polarization
\cite{Ast2007,Premper2007,Meier2008}. Similar RB-type quantum well states with significant splitting have been also observed to coexist with TI
surface states on several TI compounds where adatoms induce a strong band bending \cite{Bianchi2011,King2011,Wray2011,Benia2011,Zhu2011}.
In the field of TI, Fu has recently performed a $\mathbf{k}\cdot \mathbf{p}$ a perturbation analysis and proposed a higher order
correction to H$_{\textmd{TI}}$ which can explain the anisotropy observed in the surface states of Bi$_{2}$Te$_{3}$
\cite{Fu2009}. In this case, the extra parameter captures the deviation of their momentum distributions from perfect circles (i.e. the warping of the surface states). 

The purpose of this work is twofold. On one hand, we provide a phenomenological description which verifies that the results of $\mathbf{k}\cdot \mathbf{p}$ theory can be applied not only to TI 
but also to RB materials of the same symmetry (R3$\bar{\textmd{m}}$). The $\mathbf{k}\cdot \mathbf{p}$ Hamiltonian can be used for a common description
of RB systems whose band structure follows Eq. (\ref{EQ1}) and of those where it deviates significantly from the predictions of the classical RB effect. Band structure changes
in RB materials can be accurately captured by the continuous tuning of a dimensionless material-dependent parameter which quantifies the interplay of SO and anisotropy effects. 
On the other hand, using the same analysis, we investigate the possibility of using this parameter to discriminate the spin polarized surface states of compounds with non-trivial 
bulk topologies from those exhibiting a RB-type dispersion. Our results have been used to model the surface band structure of real systems with remarkable accuracy 
around the center of the surface Brillouin zone (SBZ). 

\section{II. Model and parameters}

As developed in Ref. \onlinecite{Fu2009}, the $\mathbf{k}\cdot \mathbf{p}$ Hamiltonian has to respect the symmetry class of the studied materials. 
Momentum and spin must be invariant under the necessary symmetry operations. Therefore, for the
R3$\bar{\textmd{m}}$ group the system Hamiltonian must take the
following form up to third order terms:
\begin{equation}
\textmd{H}=\textmd{H}_{\textmd{NFE-SO}}+\frac{c}{2}((k_{\textmd{x}}+i
k_{\textmd{y}})^{3}+(k_{\textmd{x}}-i
k_{\textmd{y}})^{3})\boldsymbol\sigma_{\textmd{z}}
\label{EQ2}
\end{equation} 

The cubic term introduces warping effects on the otherwise circular in-plane momentum distributions.
Deviations of the resulting dispersion from a single NFE parabola due to
SO coupling and hexagonal warping can be quantified by
corresponding wavevector or energy scales. The characteristic
wavevector scale for the isotropic RB effect, $k_{1}=b\cdot
m_{\textmd{s}}$ determines the $k$-shift from the parent
spin degenerate band, while the energy offset $E_{1}=(b^{2}\cdot
m_{\textmd{s}})/2$ marks the energy position of the band's
extremum. Namely, for $m^{*}>0$ the split bands have a minimum at $E=-|E_{1}|$, while for $m^{*}<0$ they have
a maximum at $E=|E_{1}|$ \cite{Ast2007,Cercellier2006}. Using dimensional
analysis, one can define a characteristic wavevector and energy
scale introduced by the hexagonal warping on a NFE 
parabola. $k_{2}=1/(3c\cdot m_{\textmd{s}})$ and 
$E_{2}=1/(54c^{2}\cdot m_{\textmd{s}}^{3})$ mark the
$k$-position and energy of the local extremum for the band dispersion normal
to a mirror plane. For $m^{*}>0$ the band has a local maximum at energy $E=|E_{2}|$, 
while for $m^{*}<0$ the band has a local minimum at energy $E=-|E_{2}|$.
A sketch of the effect of different
parameters on a model band structure is given in Fig. \ref{fig3}(b). One should bear in mind that
$k_{1}$ ($k_{2}$) and $E_{1}$ ($E_{2}$) are defined for a negligible value of
$c$ ($b$). When neither $b$ nor $c$ is small, the momentum and energy values of the RB minimum and the local maximum are only approximately given
by $k_{1}$, $k_{2}$, $E_{1}$ and $E_{2}$. Therefore, in Fig. \ref{fig3}(b), $k_{1}^{*}$, $k_{2}^{*}$, $E_{1}^{*}$ and $E_{2}^{*}$ have been used to denote the
corresponding momentum and energy positions of the extrema. In the following analysis, we will use $k_{1}$, $k_{2}$, $E_{1}$ and $E_{2}$ as the fundamental momentum and energy scales. 
The deviation of the "asterisk" values with respect to the corresponding fundamental parameters increases with the
intermixing of the Hamiltonian terms preceded by $b$ and $c$.

$E_{1}$ and $E_{2}$ are competing energy scales, since the former
favors the RB minimum for {\bf $m^{*}>0$} and the latter a local maximum in the
dispersion of the outer spin polarized branch normal to the mirror planes [see Figs. \ref{fig3}(b) and \ref{fig3}(c)]. The
characteristic wavevector scale $a=\sqrt{b/c}$ introduced by Fu is
equal to $\sqrt{k_{1}\cdot3k_{2}}$ and might be alternatively used to characterize the effects of
RB coupling and hexagonal warping \cite{Fu2009}. As noted by the author, that study discarded the effect of the effective mass
because it does not contribute significantly to the shape of the constant-energy (CE) contours.
Here, however, we will not limit our conclusions to a single material. Our aim is to comment on the possibility of a generic classification scheme by
comparing spin polarized surface states with different $m_{s}$, $b$ and $c$. For this reason, one has to follow the modifications
of the complete dispersion around $\overline{\Gamma}$ and the effective mass
is one of the key parameters. In sections III and IV
we provide evidence that the ratio $R=k_{1}/k_{2}$ (called
effective anisotropy ratio hereafter) gives us the opportunity to
describe the splitting and warping effects using a single
dimensionless constant. 2D electronic states with the same effective anisotropy ratio exhibit identical band topologies irrespective of any differences
in the values of $m_{s}$, $b$ and $c$. $R$
is also related to the ratio of the characteristic
energies $\big(R \propto \sqrt{\frac{E1}{E2}}\big)$.

In order to solve the model of Eq. (\ref{EQ2}), we used a
representation where the basis states are
$|\vec{k}\uparrow\rangle$ and $|\vec{k}\downarrow\rangle$, and spin
projections refer to the $z$ axis. If $c=0$, the corresponding `spin-up' and `spin-down' eigenstates refer to a quantization axis $\mathbf{e}_{\theta}=\mathbf{e}_z\times\mathbf{k}/k$, which is 
always perpendicular to $\mathbf{k}$, i.e. along the tangential direction. The corresponding states are 
in this case 100\% in-plane polarized, with a purely tangential spin polarization $\mathbf{P}$ -- opposite on the two branches -- rotating around $\overline{\Gamma}$.
For non-zero $c$ the quantization axis deviates from the tangential direction resulting in non-negligible radial ($\mathbf{P_{\textmd{rad}}}$) and out-of-plane ($\mathbf{P_{\textmd{z}}}$) spin polarization components.
Experimental and theoretical studies have nevertheless proven that the in-plane rotation of the spins is preserved; i.e. despite the presence of $\mathbf{P_{\textmd{rad}}}$ and $\mathbf{P_{\textmd{z}}}$,
the two eigenstates always refer to states with opposite signs of tangential spin polarization, while the total in-plane component is always tangent to the warped CE contour \cite{Premper2007,Ast2007,Meier2008}.
 \begin{figure}
  \centering
  \includegraphics[width = 7.3 cm]{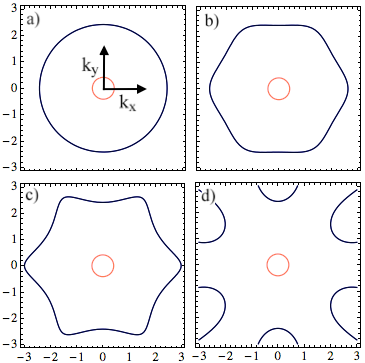}
  \caption{(color online)
Model CE contours at $E=E_{1}$ for RB systems of
R3$\bar{\textmd{m}}$ symmetry characterized by different
effective anisotropy ratios. (a) $R=0$, (b) $R=0.225$, (c)
$R=0.292$, (d) $R=0.45$. Blue (black) and red (grey) colors denote opposite
signs of tangential spin polarization. The mirror plane is along
$k_{\textmd{y}}$. The axes are scaled in units of $k_{1}$.}
\label{fig1}
\end{figure}

\section{III. Qualitative results}

\begin{figure}
  \centering
  \includegraphics[width = 7.3 cm]{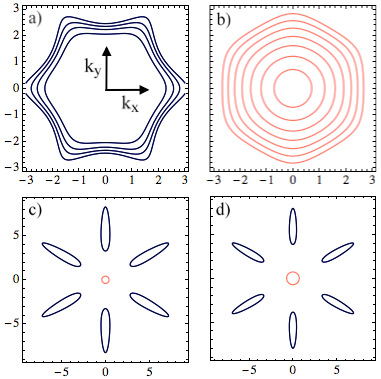}
  \caption{(color online)
[(a), (b)] Model CE contours at various energy values for (a) the outer ($E=0.1-1.1E_{1}$)
and (b) inner ($E=0.5-3.5E_{2}$) branches of the system described in Fig. \ref{fig1}(c). The contours expand
at higher energies where their sixfold symmetry becomes more pronounced.
The energy scales of (a) and (b) differ by about a factor of 10. 
[(c), (d)] The momentum distributions of Fig. \ref{fig1}(d) are
extended to larger wavevectors by using a k-quadratic
correction for $b$. The results are plotted for $E=E_{1}$ (c)
and $E=2E_{1}$ (d) and resemble the measured Fermi surface of
Bi(111) \cite{Ast2001,Koroteev2004} and Sb(111)
\cite{Sugawara2006}. Blue (black) and red (grey) colors denote opposite values of
tangential spin polarization. The mirror plane is along
$k_{\textmd{y}}$. The axes are scaled in units of $k_{1}$.}
\label{fig2}
\end{figure}

Fig. \ref{fig1} summarizes our results for an increasing
effective anisotropy ratio from (a) to (d). CE contours have
been determined from the eigenvalues of Eq. (\ref{EQ2}). The results are
plotted in units of $k_{1}$, while the energy of the contours
is equal to $E_{1}$. The image sequence clearly shows that
there is a continuous transition from the isotropic RB effect (a)
to the anisotropic case (c). The outer contour is affected by the
R3$\bar{\textmd{m}}$ symmetry of the surface and develops
hexagonal warping due to the combined effect of the threefold axis
and the time-reversal symmetry. The inner state also experiences the
in-plane symmetry of the surface as it is reflected at higher
energies [Fig. \ref{fig2}(b)]. Figs. \ref{fig2}(a) and \ref{fig2}(b) evidence that the warping for
the outer (inner) contour has a maximum positive (negative) value
normal to the mirror plane (i.e. along $k_{\textmd{x}}$), while it vanishes
along a mirror plane of the system ($k_{\textmd{y}}$). The shape of the CE contours
follows the symmetry requirements imposed by Eq. (\ref{EQ2}).
The relative in-plane orientation of the inner and outer
momentum distributions with respect to the mirror planes is in full agreement with the
conclusions of a NFE model for an anisotropic RB effect on a 2DEG with R3$\bar{\textmd{m}}$ symmetry \cite{Premper2007}.
\begin{figure}
  \centering
  \includegraphics[width = 8.7 cm]{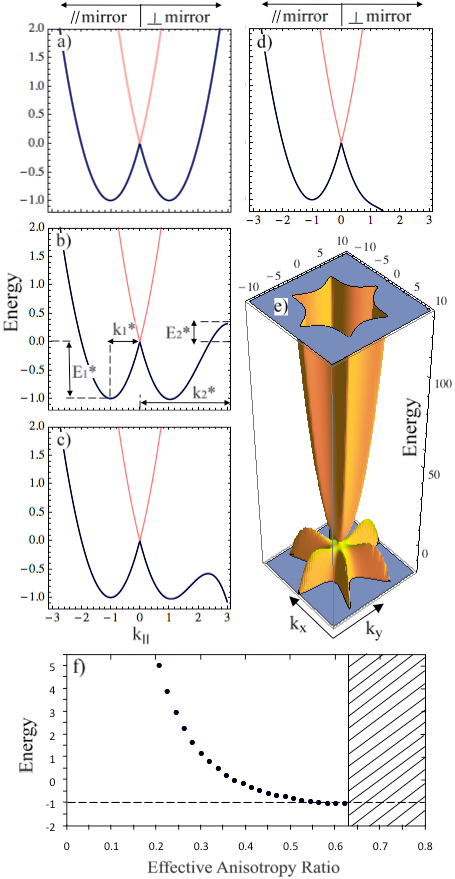}
  \caption{(color online)
[(a)-(d)] Model band dispersion for 2D states of
R3$\bar{\textmd{m}}$ symmetry characterized by different
effective anisotropy ratios. The dispersion is plotted along and normal to a mirror plane. (a) $R=0$,
(b) $R=0.349$, (c) $R=0.450$, (d) $R=0.788$. (e) The band structure
of (d) is extended to larger wavevectors by including a
$k$-quadratic correction for $b$. The parameters introduced in (b) are explained in the text. (f) The energy value of the local band maximum as a function of $R$.
The hatched area denotes the effective anisotropy ratios which are higher than $R_{\textmd{crit}}$ (see text). Blue (black) and
red (grey) colors in the band dispersion denote opposite signs of tangential spin polarization.
The mirror plane is along $k_{\textmd{y}}$. The axes are scaled in
units of $k_{1}$ and $E_{1}$, while $E=0$ determines the energy position of the spin degeneracy point.} \label{fig3}
\end{figure}
\begin{figure*}
  \centering
  \includegraphics[width = 17.9 cm]{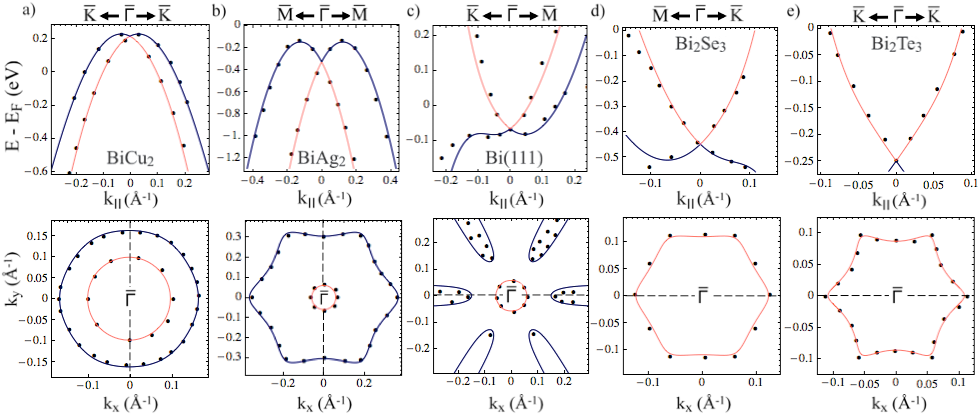}
  \caption{(color online)
Model band dispersion (upper panel) and Fermi surface contours (lower panel)
with fit parameters which match published data (black dots) on real systems with
R3$\bar{\textmd{m}}$ symmetry. Black points have been digitally acquired from the Refs. mentioned in the last column
of Table 1. All parameter values of our model are summarized in Table 1. (a) BiCu$_{2}$/Cu(111) $sp_{z}$ surface state.
(b) BiAg$_{2}$/Ag(111) $sp_{z}$ surface state: The bottom panel refers
to $E_{\textmd{B}}=-0.55$ eV. Due to the lack of high-resolution experimental CE contours, black dots in the bottom panel
refer to results from first-principles calculations \cite{Ast2007}. Before the fitting, the first-principles CE contours have been expanded to match the
corresponding experimental data [i.e. the upper panel of Fig. \ref{fig4}(b) and Refs. \onlinecite{Ast2007,Bentmann2009}]. (c) Bi(111) surface state. (d) Cu$_{0.05}$Bi$_{2}$Se$_{3}$
surface state: The experimental dispersion along $\overline{\Gamma\textmd{K}}$
(top) has been obtained by Ref. \onlinecite{Wray2010} assuming a spherical
symmetry up to $E_{\textmd{B}}=-0.2$ eV. The model results are compared
to theoretical predictions (bottom). (e) (Bi$_{1-\delta}$Sn$_{\delta}$)$_{2}$Te$_{3}$ surface state ($\delta=0.67\%$):
Corresponding fit parameters are in agreement with Ref. \onlinecite{Fu2009}.
In the bottom panel, the SBZ of the metallic surface alloys with a $\sqrt{3}\times\sqrt{3}$R$30^{\textmd{o}}$
surface reconstruction [i.e. (a), (b)] is rotated by $30^{\textmd{o}}$ with respect to the other materials [i.e. (c)-(e)].
The dashed line denotes a mirror plane dictated by Eq. (\ref{EQ2}) and it is along the $\overline{\Gamma\textmd{M}}$
direction of the SBZ. Blue (black) and red (grey) colors denote opposite
signs of tangential spin polarization.} \label{fig4}
\end{figure*}

For even larger $R$ values the outer constant energy contour is not closed
around the $\overline{\Gamma}$ point [Fig. \ref{fig1}(d)]. In order
to access its shape at larger wavevectors, one may renormalize the
value of $b$ using a recently proposed $k$-quadratic
correction [i.e. $b_{k}=b(1+\alpha k^{2})$] \cite{Fu2009}. Fig. \ref{fig2}(c) corresponds to the
anisotropy ratio of Fig. \ref{fig1}(d) with $\alpha=2$. The outer state gives
rise to six elliptical contours which shrink at higher energies
while the inner spin split branch expands [Fig. \ref{fig2}(d)]. These
contours are reminiscent of the Fermi surfaces of Bi(111), Sb(111) and the
TI compound Bi$_{\textmd{1-x}}$Sb$_{\textmd{x}}$\cite{Ast2001,Ast2002,Koroteev2004,Sugawara2006,HsiehScience2009,HsiehNatLett2008}.
All of these systems are known to have a R3$\bar{\textmd{m}}$
symmetry and exhibit two spin polarized branches. In each case,
the outer branch does not form a closed contour around
$\overline{\Gamma}$ but rather the so-called teardrop contours along
the six equivalent mirror planes of the system. The teardrop states possess an unconventional
spin polarization whose in-plane component is captured by the results presented
in Fig. \ref{fig2} \cite{HsiehScience2009}. $\overline{\Gamma}$ is one of the time-reversal
invariant momenta (TRIM) points. According to theory an electron evolving by 360$^{\textmd{o}}$ in
$k$-space around a TRIM point acquires a geometric phase (Berry's
phase) of $\pi$ \cite{Fu2007}. Figs. \ref{fig1}(d), \ref{fig2}(c) and \ref{fig2}(d) suggest that an
odd number of spin polarized contours encircle $\overline{\Gamma}$, 
which might imply a total Berry's phase of $\pi$ as a consequence of the non-trivial topological properties of the bulk. 
This is a necessary condition for the parent compounds of strong TI such as the
Sb(111) surface \cite{HsiehScience2009,Fu2007}. Although the present model
cannot follow the surface state behavior at the other TRIM points of
the surface Brillouin zone (SBZ) (i.e. $\overline{\textmd{M}}$
points), the band structure modification around $\overline{\Gamma}$ might be a hint that the value of $R$ can discriminate
between materials with topologically trivial and non-trivial band structures or consequently between pairs and an odd number of
spin-polarized surface states around the TRIM points. Quantitative remarks and caveats about this description will be
given in the following.

Fig. \ref{fig3} compares the band dispersion for different effective
anisotropy ratios and $m^{*}>0$. Along a mirror plane direction, the band structure
remains unchanged because the warping term of the Hamiltonian
is odd and has no effect. Interesting changes can, however, be observed along the
perpendicular direction: for increasing $R$,
the dispersion becomes more
anisotropic resembling the findings of Ref.
\onlinecite{Premper2007}. From (a) to (c), the outer branch continuously bends at lower energy values and
develops a local maximum [e.g. in Fig. \ref{fig3} (c)]. When the values of the fundamental parameters $E_{2}$
and $E_{1}$ become comparable, the RB minimum becomes harder to define until it is finally not observed [Fig. \ref{fig3}(d)]. In order
to access the band structure at larger wavevectors, we have plotted Fig. \ref{fig3}(d)
by using the quadratic correction for $b$. The results are presented in Fig. \ref{fig3}(e). 
The dispersion is modified in a similar way
along both high-symmetry directions by bending the outer branch at even lower energy values. The effects are more
visible at large values of $k$. As a result, the dispersion
resembles a warped cone as proposed by Fu \cite{Fu2009}.
Under the same descriptive analysis, the shape of the electronic dispersion follows the suggestions
of the electronic contours sequence presented in Fig. \ref{fig1}. 2D electronic states
similar to those encountered in an anisotropic RB NFE gas may be considered as an intermediate
case between the isotropic RB effect and surface states with a cone-like behavior
accompanied by strong hexagonal warping as observed in various TI compounds \cite{HsiehScience2009,Fu2009,Wray2010}.

Fig. \ref{fig3}(f) presents a summary of the previous observations. It illustrates, 
for non-negligible $b$ and $c$, the energy position of the local band maximum in a direction perpendicular to the mirror plane, as a
function of the effective anisotropy ratio. For increasing $R$, the energy position of the maximum approaches the Rashba minimum at a decreasing rate ($E\simeq-E_{1}$, due to 
the deviation of the fundamental parameters from the "asterisk"
values as explained in section II). This is the lowest value it can have because no local maximum is observed for higher $R$ values [Fig. \ref{fig3}(d)].
The energy value of the local maximum marks the transition from two closed concentric contours around $\overline{\Gamma}$ [e.g. Fig. \ref{fig1}(c)] to one closed contour and six additional 
elliptical pockets [e.g. Fig. \ref{fig1}(d)]. Therefore, the actual merit of Fig. \ref{fig3}(f) is that it determines the lower value of $R$ ($R_{\textmd{crit}}\simeq0.63$) for which
the spin polarized states always (i.e. at all energy values) exhibit an odd number of closed contours around $\overline{\Gamma}$.
\begin{table*}[!t]
\caption{Fit parameters of Eq. (\ref{EQ1}) used to simulate (Fig. \ref{fig4}) the topography
of different surface states with R3$\bar{\textmd{m}}$ symmetry. Regular (bold) 
fonts denote surface states of RB (TI) materials. Parameter values
follow the experimental dispersion published in the corresponding Refs. The energy value of the spin degeneracy point is set to zero.}
\begin{tabular}{c  c  c  c  c  c  c  c  c  c  c}
  \hline \hline
  material & $m^{*}(m_{\textmd{e}})$ & $b(\textmd{eV} \textmd{\AA})$ & $c(\textmd{eV} \textmd{\AA}^{3})$ & $k_{1}(\textmd{\AA}^{-1}$) & $k_{2}(\textmd{\AA}^{-1})$\; & $R$ & $E_{1}(\textmd{eV})$ & $E_{2}(\textmd{eV})$ & $E_{\textmd{F}}(\textmd{eV})$ & \;Refs\\
  \hline
  Au(111)  &  \;\;\;\;0.27\;\;\;\;\; &  \;\;\;\;\;0.33\;\;\;\;\;  &  \;\;\;\;\;0\;\;\;\;\;  &  \;\;\;\;\;0.012\;\;\;\;\;  &  \;\;\;\;\;$\infty$\;\;\;\;\;  &  \;\;\;0\;\;\;  &  \;\;\;\;\;0.002\;\;\;\;\;  &  \;$\infty$\;\;  &  \;\;0.475\;\;  &  \;\cite{LaShell1996,Cercellier2006}\\
  Bi/Cu(111)  &  \;\;\;-0.29\;\;\;\;\;  &  \;\;\;\;\;0.85\;\;\;\;\; &  \;\;\;\;\;12\;\;\;\;\;  &  \;\;\;\;-0.032\;\;\;\;\;  &  \;\;\;\;-0.741\;\;\;\;\;  &  \;\;\;0.043\;\;\;  &  \;\;\;\;-0.014\;\;\;\;\;  &  -2.439\;\;  &  \;-0.215\;\;  &  \;\cite{Bentmann2009,Moreschini2009}\\
  Bi/Ag(111)  &  \;\;\;-0.32\;\;\;\;\;  &  \;\;\;\;\;2.95\;\;\;\;\; &  \;\;\;\;\;18\;\;\;\;\;  &  \;\;\;\;-0.122\;\;\;\;\;  &  \;\;\;\;-0.446\;\;\;\;\;  &  \;\;\;0.274\;\;\;  &  \;\;\;\;-0.181\;\;\;\;\;  &  -0.800\;\;  &  \;\;N/A\;\;  & \cite{Ast2007,Bentmann2009,Meier2008}\\
  Bi(111)  &  \;\;\;\;0.50\;\;\;\;\;  &  \;\;\;\;\;0.70\;\;\;\;\;  &  \;\;\;\;\;58\;\;\;\;\;  &  \;\;\;\;\;0.046\;\;\;\;\;  &  \;\;\;\;\;0.088\;\;\;\;\;  &  \;\;\;0.515\;\;\;  &  \;\;\;\;\;0.016\;\;\;\;\;  &  \;0.020\;\;  &  \;\;0.067\;\;  &  \;\cite{Ast2001,Koroteev2004}\\ \hline
  \textbf{Cu$_{0.05}$Bi$_{2}$Se$_{3}$}  &  \;\;\;\;0.26\;\;\;\;\;  &  \;\;\;\;\;1.90\;\;\;\;\; &  \;\;\;\;\;140\;\;\;\;\;  &  \;\;\;\;\;0.067\;\;\;\;\;  &  \;\;\;\;\;0.068\;\;\;\;\;  &  \;\;\;0.978\;\;\;  &  \;\;\;\;\;0.063\;\;\;\;\;  &  \;0.022\;\;  &  \;\;0.45\;\;  &  \;\cite{Wray2010}\\
  \textbf{(Bi$_{1-\delta}$Sn$_{\delta}$)$_{2}$Te$_{3}$}  &  \;\;\;\;1.37\;\;\;\;\;  &  \;\;\;\;\;2.00\;\;\;\;\; &  \;\;\;\;\;230\;\;\;\;\;  &  \;\;\;\;\;0.360\;\;\;\;\;  &  \;\;\;\;\;0.008\;\;\;\;\;  &  \;\;\;44.444\;\;\;  & \;\;\;\;\;0.360\;\;\;\;\;  &  \;6$\times10^{-5}$\;\;  &  \;\;0.25\;\;  &  \;\cite{HsiehScience2009,Fu2009}\\ \hline \hline
\end{tabular}
\end{table*}

\section{IV. Quantitative results}

We will now use the results of the previous section in a quantitative analysis of published data on
real systems. Table 1
lists the effective anisotropy ratios, as well as the
characteristic wavevector and energy scales for five different
materials with R3$\bar{\textmd{m}}$ symmetry. Results of the
fits are depicted in Fig. \ref{fig4} and summarized in Table 1. The model of Eq. (\ref{EQ2}) has been solved using values of $m^{*}$ and $b$
in good agreement with those in the corresponding Refs., while $c$ is treated as the
only free parameter. 

The surface state of Au(111) is included as the paradigm of an isotropic RB splitting. 
The consequences of the anisotropy start
appearing for the spin split $sp_{z}$ surface states of
the BiCu$_{2}$/Cu(111) surface alloy which exhibits a non-zero
$R$ value. Nevertheless,
anisotropy effects are not readily seen at the Fermi surface
contours [Fig. \ref{fig4}(a) and Refs. \onlinecite{Moreschini2009,Bentmann2009}] because
the Fermi energy value is very small compared to $E_{2}$. The
corresponding $sp_{z}$ states of the BiAg$_{2}$/Ag(111)
alloy are more anisotropic. This can be quantified by a substantial
increase in the value of $R$. These surface states do not exhibit any Fermi
surface contours but the anisotropy manifests itself in the
CE maps taken at  a binding energy $E_{\textmd{B}}$ somewhat higher than the
spin degenerate point [Fig. \ref{fig4}(b)].

The effective anisotropy ratio of the Bi(111) surface states is                                                    
even higher and $E_{\textmd{F}}$ is now comparable to $E_{2}$.
The anisotropy effects are very strong on the
Fermi surface contours. They consist of a single
spin polarized state centered at $\overline{\Gamma}$ and six hole pockets \cite{Ast2001,Ast2002,Koroteev2004}, which are
reproduced by our results. As
pointed out previously, such surface CE contours have been observed in materials with topologically non-trivial band structures
\cite{HsiehScience2009,HsiehNatLett2008}. According to the tight-binding model of Ref.
\onlinecite{Koroteev2004}, the surface states of Bi(111) exhibit a complex dispersion
at larger wavevectors. Therefore, the complete bandstructure
can be revealed only by first-principles methods \cite{Koroteev2004}.
Nevertheless, our simple NFE-based results can provide valuable information 
about the dispersion in regions where the
parent band can be approximated by a simple parabola, since they indicate that the outer branch 
forms contours which do not close around $\overline{\Gamma}$.

The 3D TI compounds Bi$_{2}$Se$_{3}$ (Cu$_{x}$Bi$_{2}$Se$_{3}$ with $x=0.05$)
and Bi$_{2}$Te$_{3}$ (Sn-doped 0.27\%) have been included as examples of materials
with R3$\bar{\textmd{m}}$ symmetry and high effective anisotropy ratios. We have chosen these
members of the TI family due to availability of high quality data which readily shows warping effects on the band
dispersion and the CE contours \cite{Wray2010,Chen2009}. The fit parameters reproduce very well the experimental results
[Figs. \ref{fig4}(d) top and \ref{fig4}(e)] or the theoretical predictions [Fig. \ref{fig4}(d) bottom] for small $k$-values. The outer spin polarized branch forms the lower part of the
dispersion cone and there is a single spin polarized contour around the center of the SBZ, as expected for materials with a topologically non-trivial
band structure. The effective anisotropy ratio of Bi$_{2}$Te$_{3}$ is more than one order of magnitude higher
than the one of Cu-intercalated Bi$_{2}$Se$_{3}$ as revealed by the corresponding parameters and reflected in the blossomlike shape of the Fermi surface 
contour. The hexagonal warping term $c$ used for Bi$_{2}$Te$_{3}$ is in close agreement with the one proposed by Fu in his recent study
\cite{Fu2009}. 

The prediction of an unmodified RB dispersion along the mirror plane direction of the 3D TI materials (i.e. not taking into account
a quadratic correction in the value of $b$) may seem at a first glance in striking contrast with the well-known cone-like dispersion of their surface states. 
Nevertheless, one should bear in mind that at energies slightly lower than the Dirac spin degeneracy point the surface states of TI materials merge into the bulk valence band continuum, thereby 
obscuring any clear experimental identification \cite{Wray2010,Chen2009}. Such a difference in the dispersion between the mirror plane 
direction and its perpendicular has been clearly identified by \textit{ab initio} calculations on various compounds of the TI family \cite{Zhang2009}.

\section{V. Conclusions}

Warped in-plane momentum distributions like those depicted in Fig. \ref{fig4} are a signature of a $k$-dependent out-of-plane
spin polarization component $\mathbf{P_{\textmd{z}}}$ \cite{Premper2007,Fu2009,Meier2008,Ast2007}.
According to previous theoretical studies and experimental results, $\mathbf{P_{\textmd{z}}}$ vanishes along
the three equivalent surface mirror planes \cite{Premper2007,Fu2009,Meier2008}, one of which is indicated by the dashed lines in Fig. \ref{fig4}, while the other two
are rotated by $\pm120^{\textmd{o}}$. The value of $P_{z}$ follows a threefold symmetry
determining the overall symmetry of the total spin polarization vector under the requirements of Eq. (\ref{EQ2}).
Moreover, the orientation of the inner and outer CE contours is a fingerprint for the symmetry of the $\mathbf{P_{\textmd{z}}}$
vector \cite{Premper2007,Meier2008}. Therefore, the results presented in the bottom panel of Fig. \ref{fig4} can qualitatively capture the
$k$-dependent out-of-plane rotation of the total spin polarization vector.

As a tradeoff for its simplicity, the Hamiltonian proposed by Fu [i.e. Eq. (\ref{EQ2})] has a few shortcomings.
Since it is derived by $\mathbf{k}\cdot \mathbf{p}$ perturbation theory, it is valid at wavevector values around $\overline{\Gamma}$. 
In order to satisfy this need we did not extend our results to other high-symmetry points of the SBZ. 
One should bear in mind that, similar to Ref. \onlinecite{Fu2009}, we used scaled axes in Figs. \ref{fig1}-\ref{fig3}. 
Scaling is necessary to maintain a universal character of the trends we have addressed. 
Nevertheless, the validity of the corresponding results at points far from the SBZ center depends on the absolute $k$-distance from zero. 
In this frame, Fig. \ref{fig4} proves that the conclusions from the scaled-axes Figures can indeed be applied to real systems. 
Another shortcoming of the model is related to the passage from anisotropic RB-type 2D electronic states to 
surface states exhibiting a dispersion similar to those of TI compounds. It is not straightforward to claim a topological change because
our results are limited to the surface states and they do not consider 
the bulk valence and conduction bands of the described compounds. However, the energy 
position of the bulk bands continuum is crucial to determine whether there is a partner switching in the spin polarized surface state branches within the bulk gap and between two TRIM points \cite{Fu2007}.
The topological differences between similar compounds may depend on bulk band characteristics such as a critical bulk band inversion which discriminates the
trivial band structure of Bi(111) from the non-trivial band structure of Sb(111) \cite{FuPRB2007}. Although the described model cannot capture such
bulk effects, the effective anisotropy ratio is a measure of how fast the surface state dispersion deviates from the isotropic case. 
Therefore, it may determine the minimum size of the bulk gap around $\overline{\Gamma}$ which would allow a partner switching to occur. 
Although Eq. (\ref{EQ2}) has already been used to describe the surface states of Bi$_{2}$Te$_{3}$ \cite{Fu2009}, the derived dispersion
along the direction which connects two TRIM points remains unchanged in the absence of the $k$-quadratic correction in the value of $b$.
As a matter of fact, the best indication for a possible topological change within the model is the predicted change in the total number of closed spin polarized contours around $\overline{\Gamma}$. 
According to Fig. \ref{fig3}(f), there will always be an odd number of closed contours for $R\geq0.63$. In any case, since a purely two-dimensional model is not
enough to capture the complete band structure of a TI compound, our results on this topic rest as
a descriptive analysis of the alleged topological change. Final predictions should be made
after determining the surface state behavior at different TRIM points which cannot be captured by Eq. (\ref{EQ2}). 

Nevertheless, the value of the present work is not limited to a purely phenomenological analysis
of the consequences of the aforementioned topological change reflected on the surface states. The description of the RB effect using
the Fu Hamiltonian [Eq. (\ref{EQ2})], although overlooked in the past RB literature, offers major advantages. As explicitly demonstrated by
our qualitative and quantitative arguments, it presents a simple but instructive way to capture the effects of anisotropy 
on the surface band structure. Continuous tuning of the effective anisotropy ratio permits a unified description of different RB systems with complex topologies
without resorting to relativistic DFT calculations and deviating from the conventional RB description. The same Hamiltonian [i.e. Eq. (\ref{EQ2})] 
can capture the band structure of (a) the classical RB effect as observed on Au(111)
and was originally accessed by the isotropic RB Hamiltonian [i.e. Eq. (\ref{EQ1})] \cite{LaShell1996}, (b) the "weakly" anisotropic RB systems such as the metallic surface alloys which have been
previously described by DFT calculations or by an NFE model with an \textit{ad hoc} in-plane potential gradient \cite{Ast2007,Premper2007}, and (c) "strongly" anisotropic
cases which do not show a pair of closed contours around the spin degeneracy points and have been previously described only by relativistic first-principles calculations \cite{Koroteev2004}. Differences between the corresponding RB surface states are accurately captured by the effective anisotropy ratio $R$, which can be determined either from an experimental CE contour or from
the band dispersion perpendicular to the mirror plane direction. Alternatively, the simple $R$ parameter can be determined by fitting relativistic first-principles calculations.          
                                                                                                                                                                                                                                                                                                                                                                
Can we attribute a physical meaning to the effective anisotropy ratio? 
As already mentioned, $R$ quantifies the warping effects on the surface state dispersion, which are always 
accompanied by a non-zero out-of-plane spin polarization component. The latter is intuitively linked to an in-plane electronic potential gradient. 
Premper et al. introduced explicitly such an in-plane gradient in the standard RB model and found that it can both influence the splitting 
size and alter the circular shape of the surface state contours \cite{Premper2007}. There are however alternative views of the same phenomenon.
Bihlmayer et al. proposed that it is the asymmetry of the surface state wavefunction 
which drives the size and anisotropy of the splitting \cite{Bihlmayer2006}. 
Recent first-principles calculations suggested that the asymmetric charge distribution can arise from the unquenched local orbital angular momentum in the presence of SO coupling \cite{Park2011}. 
Because of their computational complexity, 
such approaches face practical limitations, and the
effective anisotropy ratio $R$ can be considered as an alternative easy-to-use effective parameter, which 
quantifies the warping effects induced to surface states of R3$\bar{\textmd{m}}$ symmetry according to the predictions of the aforementioned physical models.
Therefore, $R$ is intimately related to an in-plane asymmetry of the electronic charge distribution, which in turn 
may correspond to an asymmetric surface state wavefunction or again to an anisotropic quenching of the local orbital angular momentum.

\section{VI. Summary}

We presented a phenomenological description of the similarities and differences between the surface states of RB and
TI compounds by continuously tuning an anisotropy-related parameter. We showed that
the $\mathbf{k}\cdot \mathbf{p}$ Hamiltonian proposed by Fu \cite{Fu2009} can be
used not only to describe the surface states of TI compounds but also as a
tool to quantify anisotropy effects in RB systems. We propose a
general classification scheme for 2D spin polarized states of R3$\bar{\textmd{m}}$ symmetry. 
Materials which exhibit an isotropic RB splitting,
and 3D topological insulators with a single Dirac cone on their
surface can be considered as the two opposite ends of our
classification. Systems with an anisotropic RB splitting naturally fall 
into place as an intermediate category. 
This simple approach suggests a unified description of the RB surface states on different materials which 
exhibit complex spin polarized contours without
deviating from the simplicity of the conventional RB Hamiltonian. We hope that it will motivate more fundamental efforts to compute the in-plane electronic charge distribution and 
the momentum and energy dependence of the spin polarization vector.
 
\begin{acknowledgments}
We are grateful to A. Crepaldi and S. Pons for fruitful discussions and their helpful comments.
This research was supported by the Swiss NSF and the NCCR MaNEP.
\end{acknowledgments}

\end{document}